\documentclass[reprint, amsmath,amssymb, aps,]{revtex4-2}

\usepackage{graphicx}
\usepackage{dcolumn}
\usepackage{bm}

\begin{document}

\preprint{APS/123-QED}

\title{Following enhanced Sm spin projection in Gd$_x$Sm$_{1-x}$N} 

\author{J. D. Miller}
 \email{jackson.miller@vuw.ac.nz}
\author{J. F. McNulty}
\author{B. J. Ruck}
\affiliation{%
 MacDiarmid Institute for Advanced Materials and Nanotechnology, \\ School of Chemical and Physical Sciences, \\ Victoria University of Wellington, P.O. Box 600, Wellington 6140, New Zealand
}%
\author{M. Al Khalfioui}
\author{S. V\'{e}zian}
\affiliation{Universit\'{e} C\^{o}te d'Azur, CNRS-CRHEA, Valbonne 06560, France}

\author{M. Suzuki}
\altaffiliation[Present address: ]{School of Engineering, Kwansei Gakuin University, 2-1 Gakuen, Sanda, Hyogo 669-1337, Japan}
\author{H. Osawa}
\author{N. Kawamura}
\affiliation{Japan Synchrotron Radiation Research Institute (JASRI), 1-1-1 Kouto, Sayo, Hyogo, 679-5198, Japan
}%

\author{H. J. Trodahl}%
 \email{joe.trodahl@vuw.ac.nz}
\affiliation{%
School of Chemical and Physical Sciences, Victoria University of Wellington, \\ P.O. Box 600, Wellington 6140, New Zealand
}%

\date{\today}

\begin{abstract}
The rare-earth nitrides form a series of structurally simple $intrinsic$ ferromagnetic semiconductors, a rare class of both fundamental interest and application potential. Within the series there is a wide range of magnetic properties relating to the spin/orbit contributions to the ferromagnetic ground states. We report an x-ray magnetic circular dichroism investigation of the spin/orbit magnetic dipole alignments of Sm and Gd ions in epitaxial Gd$_x$Sm$_{1-x}$N films. The Sm spin-alignment expectation value $\langle S_{z} \rangle$ is seen to be strengthened by the Gd/Sm exchange interaction, providing guidance concerning the composition for an angular momentum compensation point (where the volume-averaged total angular momentum of a film is zero).
\end{abstract}

\maketitle

\section{Introduction}

The lanthanides are the series of the periodic table across which the $4f$ shell is sequentially filled, imbuing them with the strongest angular momentum and magnetic moments among all stable elements. The strong spin-orbit coupling within the 4$f$ shell ties the orbital moment firmly to the spin so that the net magnetic moment is of mixed spin and orbital character. Furthermore, the limited extent of the $4f$ wave function reduces the orbital angular momentum quenching that is a feature of transition metal compounds; indeed the orbital moment can dominate the net magnetic moment.

The rare-earth nitrides (RN, with R a lanthanide element) form in the very simple NaCl structure which provides a fertile ground to investigate the spin/orbit magnetism that stems from the partially full $4f$ electron levels~\cite{Natali2013}. 
The chemical similarity of the lanthanides and the well-matched lattice constants in their nitrides invites studies in the mixed-cation environments offered by multilayers and solid solutions. Multilayer structures of GdN/SmN and of GdN/NdN have already shown a striking competition between the exchange and Zeeman interactions across interfaces~\cite{McNulty2015, Anton2021,Anton2016,McNulty2019}. The present report investigates the effects of that competition among randomly-sited Gd$^{3+}$/Sm$^{3+}$ ions in the solid solution Gd$_{x}$Sm$_{1-x}$N using x-ray magnetic circular dichroism (XMCD).

The magnetism of the parent materials, GdN ($x=1$) and SmN ($x=0$) illustrates the range of magnetism in the RN series. Both are ferromagnetic, with Curie temperatures (T$_C$) of $\sim70$ and $\sim30$~K, respectively, resulting from an indirect exchange mechanism that links the $4f$ alignments via the R $5d$ and N $2p$ states \cite{Hulliger1978,Natali2013,Granville2006,Meyer2008}. The Gd$^{3+}$ ions in GdN have a half-filled $4f$ shell ([Xe]$4f^{7}$), and in the ground state all seven spins align giving a spin-number, $S~=~7/2$. The half-full shell further features zero orbital angular momentum ($L = 0$) so that the material behaves as a spin-only ferromagnet with $\langle M_{z} \rangle = 2 \mu_{B} \langle S_{z} \rangle = 7~\mu_{B}$ per Gd$^{3+}$ ion. In contrast, the Sm$^{3+}$ ions in SmN possess two fewer electrons giving an electronic configuration of [Xe]$4f^{5}$. Inter-ion exchange aligns the $4f$ spins, but opposing magnetic contributions from the spin and orbital angular momenta very nearly cancel to yield a net magnetisation of $\approx$~0.035~$\mu_B$ per Sm$^{3+}$ ion in the ferromagnetic ground state~\cite{Moon1979, Meyer2008}.  Crucially, the magnetic moment $\langle M_{z} \rangle = \mu_{B} (\langle L_{z} \rangle + 2 \langle S_{z} \rangle)$ is weakly orbital dominated, with $\langle L_{z} \rangle / 2 \langle S_{z} \rangle < -1$~\cite{Anton2013}, so that the spin contribution to the magnetic moment is opposed to an applied field. It is this opposition of the spin magnetic moment ($\mu_{S}$) to the applied field that provides a competition between the Zeeman and exchange interactions when SmN and GdN are paired in multilayers~\cite{Anton2013,Anton2021}.

\begin{figure}
     \centering
     \includegraphics[width=\linewidth]{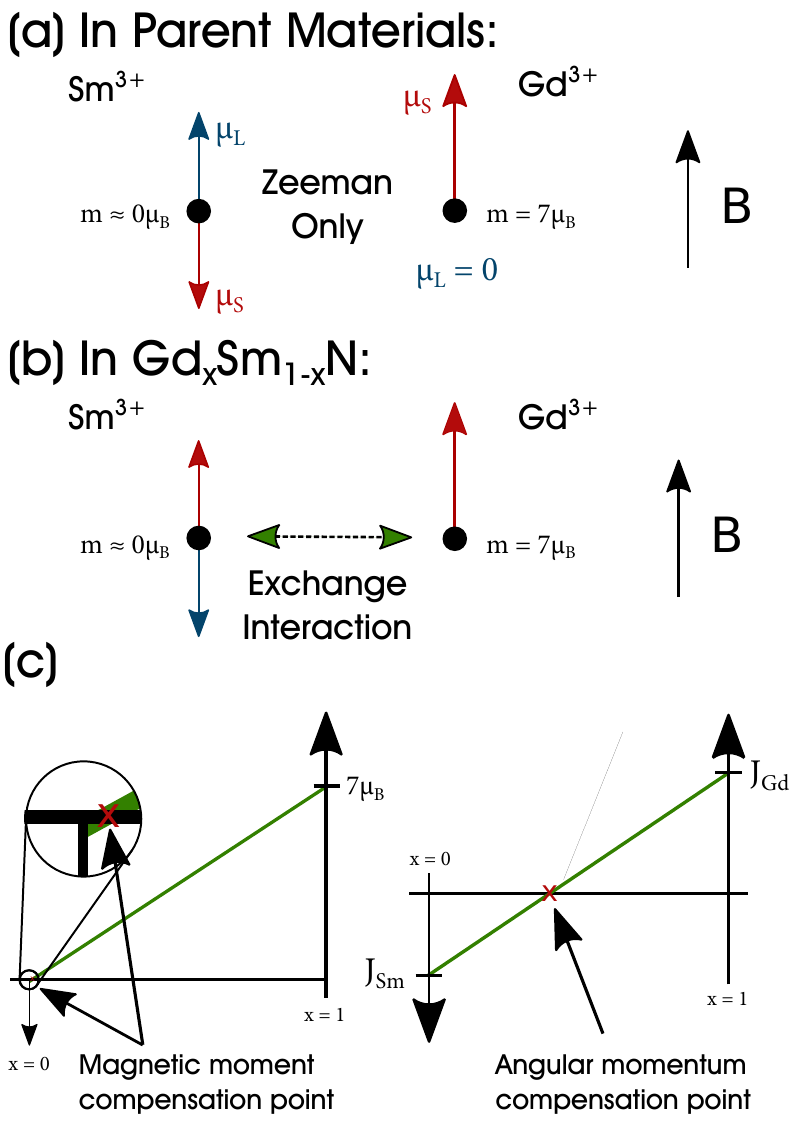}
     \caption{Schematic showing the alignment of the magnetic moments of Sm$^{3+}$ and Gd$^{3+}$ in (a) the parent materials where the Sm-Gd exchange is absent and (b) the solid solution Gd$_{x}$Sm$_{1-x}$N where the Sm-Gd exchange dominates the Zeeman interaction and aligns the spins. Panel (c) shows the net (low-temperature, volume averaged) values of the magnetisation ($\bar{M}$) and angular momentum ($\bar{J}$) as the composition ($x$) interpolates between the parent materials. This schematic refers to to the most simple model and is shown in Figure~\ref{J_x} adapted in light of our XMCD results.}
     \label{Schematic}
 \end{figure}

Similarly to multilayer structures, the Gd-Sm exchange interaction in Gd$_{x}$Sm$_{1-x}$N disrupts the tendency of the Zeeman interaction to align the Sm$^{3+}$ orbital magnetic moment ($\mu_{L}$) with an applied field. The alignment (anti-alignment) of the Gd$^{3+}$ (Sm$^{3+}$) spin magnetic moments with an applied magnetic field (Figure~\ref{Schematic}a) is in direct competition with the ferromagnetic inter-ion exchange attempting to align the Gd$^{3+}$ and Sm$^{3+}$ spins (Figure~\ref{Schematic}b). Because the Gd$^{3+}$ magnetic moment is two orders of magnitude the larger, most values of $x$ show the alignment of the the Gd$^{3+}$ spin-moment with the applied field. Yet, for extremely dilute concentrations of Gd in solution (that is, $x \approx 0$) it must be that the net Sm$^{3+}$ magnetic moments align with the applied field, rather than the Gd$^{3+}$, due to the same exchange. 





The simplest model for the magnetic state of Gd$_{x}$Sm$_{1-x}$N would treat the Sm$^{3+}$ and Gd$^{3+}$ ion moments as fixed at their values in SmN and GdN. This model suggests that under full ferromagnetic order the saturation magnetisation is linear in $x$ and for some appropriate composition there exists a net zero magnetisation, a magnetisation compensation point. Within that model the net magnetisation ($\bar{M}$) ranges from $-0.035~\mu_{B}$ to $7~\mu_{B}$ per R ion (with respect to the spin-moment direction in the solid solution) with the compensation point occurring at $x~=~0.005$ (Figure \ref{Schematic}c). The XMCD results in this paper indicate that the Sm$^{3+}$ spin-alignment does $not$ remain as adopted in SmN, but nearly doubles in Gd-rich films. 

Furthermore, the same behaviour seen for the net magnetic moment must also manifest for the net angular momentum, $\bar{J}$, meaning that under full FM order $\bar{J}$ $must$ also pass through zero as $x$ varies from $0$ to $1$, giving an angular momentum compensation point (Figure~\ref{Schematic}c). Here again the simplest model uses a linear interpolation between the ground state angular momenta of GdN and SmN, but unlike the magnetisation there is no simple measurement of the angular momenta of the parent materials. The spin-only magnetism in GdN clearly indicates that $J=S=7/2$, and Hund's rules suggest that $J=L-S=5/2$ in SmN. However, an atom-centred treatment of the magnetism of SmN based on realistic crystal-field and exchange interactions and reproducing the orbital-dominant ground state moment of $0.035~\mu_{B}$ per Sm$^{3+}$ ion yields an eigenstate of neither $L_{z}$ or $S_{z}$~\cite{McNulty2016}. Rather, it shows projections $\langle L_{z} \rangle \approx 2\hbar$ and $\langle S_{z} \rangle \approx -\hbar$, substantially smaller than the maximum possible values provided by Hund's rules ($\langle L_{z} \rangle = 5\hbar$ and $\langle S_{z} \rangle~=-5\hbar/2$, treating the spin moment direction as positive, once more). Importantly, the spin-orbit coupling is still strong enough that the spin and orbital angular momenta remain opposed. 

This description of the magnetism of the Sm$^{3+}$ ion, in which the $4f$ spins are ferromagnetically aligned in the ground state, but without their maximum potential magnitudes, accommodates the factor of $\approx$~2 increase in the Sm spin polarisation that is seen in XMCD measurements performed on GdN/SmN superlattices~\cite{McNulty2015,McNulty2016}. We thus introduce a parameter $\alpha = \langle S_{z} \rangle / (5\hbar/2)$, which indicates the expectation value of the spin-alignment $\langle S_{z} \rangle$ in the $4f$ shell of the Sm$^{3+}$ ion. Appealing to the ion-based model of \textcite{McNulty2016}, in homogeneous SmN $\alpha~\approx~0.4$, while at the SmN/GdN interface $\alpha~\approx~0.6$. One might then anticipate, as seen below, that that $\alpha$ increases with Gd concentration $x$. The simple model shown in Figure~\ref{Schematic} illustrates the dependence of the net magnetic moment and angular momentum on $x$, but assumes that $\alpha$ is independent of $x$. In the upcoming discussion we show the equivalent diagram with $\alpha(x)$ as implied by the XMCD results (Figure~\ref{J_x}).


\section{Experimental Details}

Gd$_{x}$Sm$_{1-x}$N films were grown in a Riber 32p molecular beam epitaxy (MBE) system on 100 nm thick (0001) AlN buffered Si, heated during growth to between $450$ and $700^{\circ}$~C depending on the composition. The growth rate was held to $100$~nm/h. High purity rare-earth metals were evaporated from conventional effusion cells to achieve the desired composition under nitrogen rich conditions ($P_{N_2} \approx 2.1\times10^{-5}$~Torr). The composition of the solid-solutions was measured to within $x \pm 0.04$ using X-ray fluorescence (XRF) spectroscopy. X-ray diffraction confirmed (111)-oriented FCC NaCl structure, typical of the rare-earth nitride series. 



Field in-plane SQUID magnetometry was performed in a Quantum Design Magnetic Property Measurements System (MPMS) capable of temperatures as low as 5~Kelvin and fields up to 7~Tesla. These measurements determine the temperature- and hysteretic field-dependent net magnetic moment across the para- to ferro-magnetic transition. In the present system that net magnetic moment is overwhelmingly dominated by the Gd ions, motivating support of the ionic-species resolution capability of XMCD to investigate the Sm$^{3+}$ alignment separately. That alignment relates specifically to the 4$f$-shell spin and orbital states, suggesting the use of XMCD at the M$_{4,5}$ edge, which interrogates alignment in the 4$f$ shell. However, the surface sensitivity of XAS at the $\sim 1$ keV M-edge renders it unsuitable for the present study, and we have used instead XMCD at the L$_{2,3}$ edges of Gd and Sm. The hard-x-ray lanthanide L edge measurements probed the full $\approx$ 100 nm thickness of the films.

The L edge involves excitation into the $5d$ band, coupled to the $4f$ spin alignment by the strong $5d$-$4f$ exchange interaction. Thus, we report here the results of XMCD at the Gd and Sm L$_{2,3}$ edges to probe separately the alignment on the two cations. The measurements were performed at temperatures down to $5$~K and fields up to $6$~T at the BL39XU beamline at the SPring-8 synchrotron in Japan. All XMCD was performed at $10 ^{\circ}$ from grazing incidence.


XMCD at the two edges, L$_{2,3}$, separated by the spin-orbit interaction in the $2p$ core-hole state, can in principle yield information about the spin and orbital alignments in the final $5d$ state, though the failure of the XMCD sum rules for the L$_{2,3}$ edges prevent their use in the present case~\cite{Parlebas2006}. Furthermore, the Sm L$_2$ edge in these materials is masked by magnetic EXAFS from the Gd L$_3$ edge~\cite{Schutz1996}. We thus focus the present discussion on the relatively weak L$_3$ edge of Sm and the stronger L$_2$ edge of Gd, interpreted in terms of the temperature- and composition-dependent spin alignments. 



\begin{figure}
    \centering
    \includegraphics[width=\linewidth]{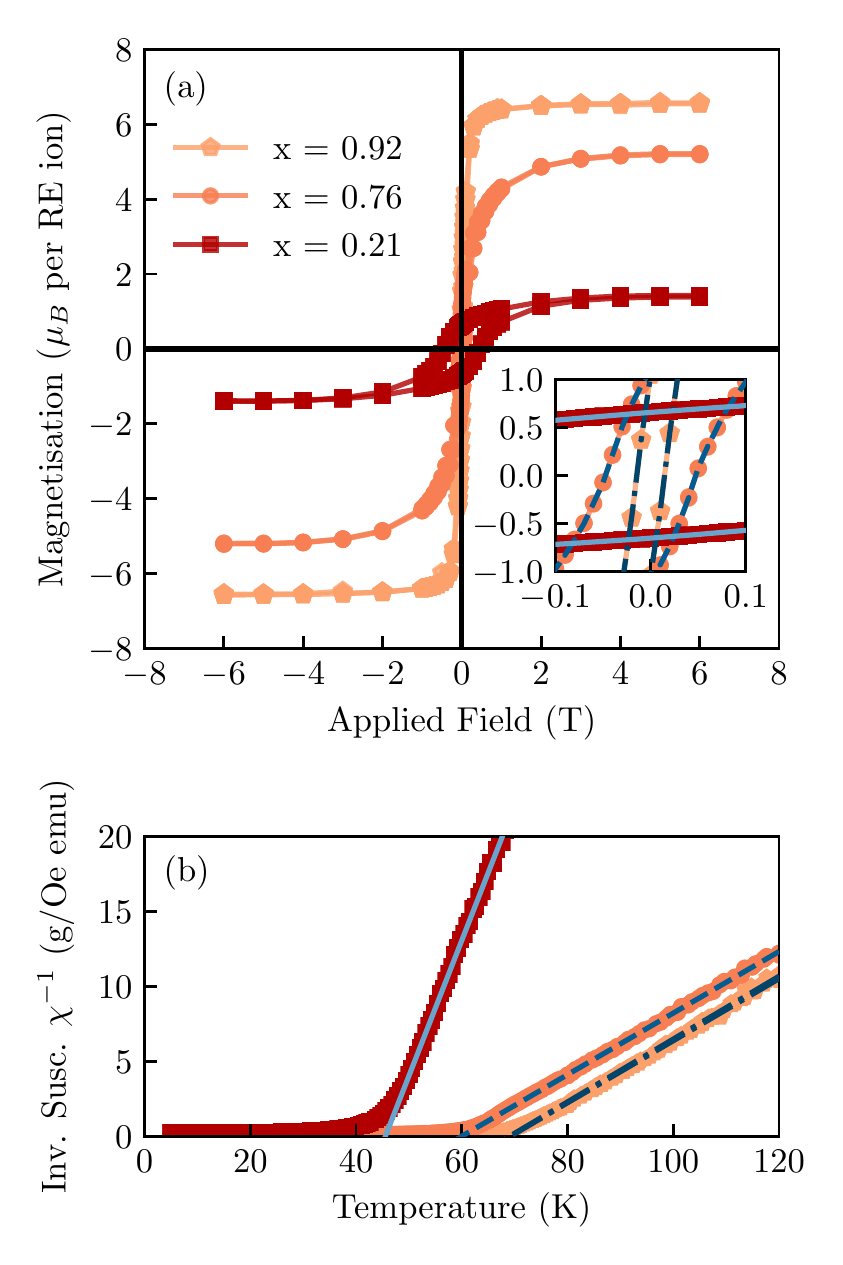}
    \caption{(a) Hysteresis loops of the three Gd$_x$Sm$_{1-x}$N films taken at $5$~K. The inset shows a zoom of the region close to zero, emphasising the different coercivities of the films. (b) The inverse susceptibility of the three films with fits included to show the different Curie temperatures extracted from the measurements.}
    \label{SQUID_Loops}
\end{figure}

\section{Results and Discussion}
\subsection{Magnetisation in the para- and ferro-magnetic phases}

To aid the interpretation of the XMCD results, we present field-cooled magnetization and hysteresis loop measurements (Figure \ref{SQUID_Loops}) on the same three films for which XMCD was performed. The temperature traces identified for each sample a single ferromagnetic transition between the 30 and 70 K Curie temperatures of SmN and GdN~\cite{Meyer2008, Granville2006, Ludbrook2009}, as seen in Figure \ref{SQUID_Loops}b. The most Gd-rich film ($x = 0.92$) has the smallest coercive and saturation fields, comparable with epitaxial GdN films. With $\approx$ 1/4 of the Gd$^{3+}$ ions replaced by non-spherically-symmetric Sm$^{3+}$ there are increases in both the coercive and saturation fields. Both trends, which are further enhanced in the most Sm-rich film, follow from the combination of (i) a weaker Zeeman interaction that results from a reduced saturation magnetisation, and (ii) a larger crystal anisotropy experienced by the $J = 5/2$ ground state of Sm$^{3+}$. All films show a full alignment above $\sim 5$~T, dictating that the spectra below were collected at the maximum available XMCD field of $6$~T.



\subsection{XMCD}

\begin{figure}
	\includegraphics[width=\linewidth]{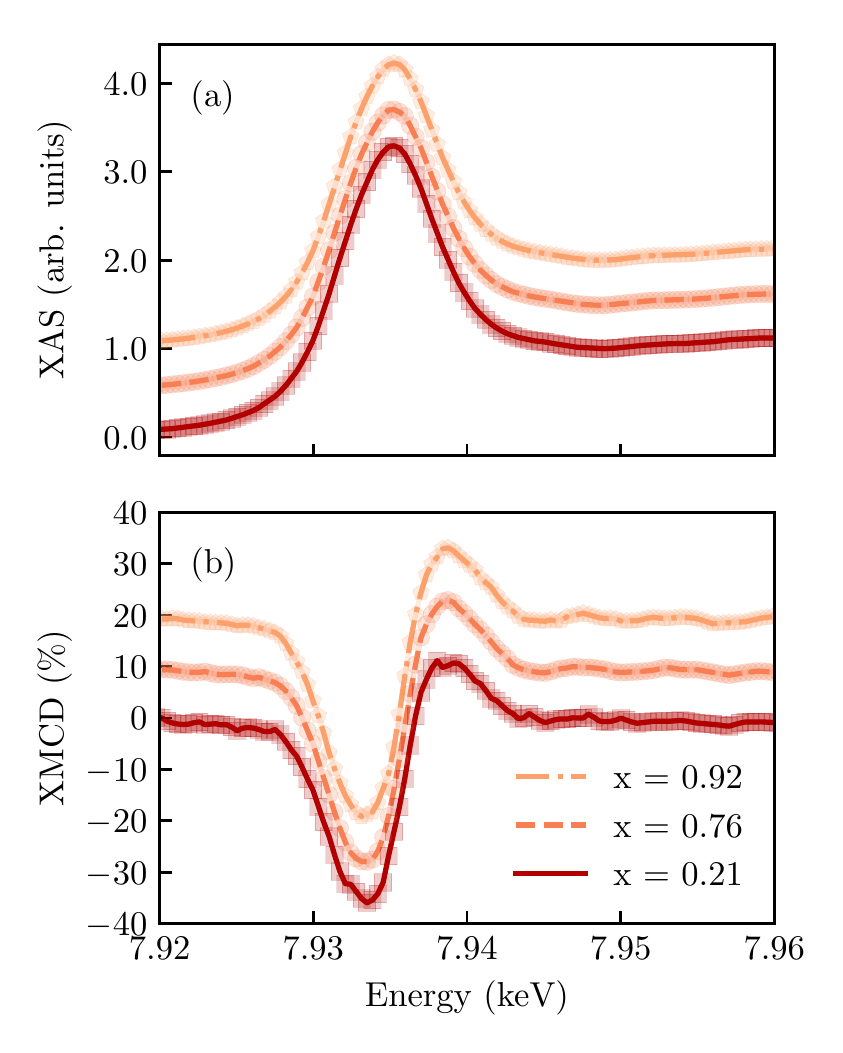}
	\caption{\small Comparison of XMCD spectra for the three Gd$_x$Sm$_{1-x}$N films at the Gd L$_{2}$ edge at $5$~K. All measurements were cooled to $5$~K in an applied field of 6~Tesla. Panel (a) shows the average of the x-ray absorption measurements for left- and right- circularly polarised light, normalised to the continuum edge jump. Panel (b) shows the XMCD, defined as the difference between the left- and right- circularly polarised absorption.}
	\label{Alloys-GdL3}
\end{figure}

The near-zero magnetic-moment of Sm in the ferromagnetic state of SmN ensures that the magnetisation of Figure \ref{SQUID_Loops} relates directly to the Gd alignment. We thus show first in Figure \ref{Alloys-GdL3} the Gd L$_2$-edge XAS (a) and XMCD (b) in the three films taken at saturation, 5 K and 6 T. A single feature dominates here, with the nearly identical magnitudes of the XMCD for different sample compositions providing a reminder that normalising the XMCD to the XAS signal causes the XMCD to reflect the \emph{alignment} of the magnetic moments rather than their concentration. The few \% reduction in the most Sm-rich film then signals that the Gd$^{3+}$ ions are less perfectly aligned in that film. This less-than-perfect alignment is discussed further where we present the temperature-dependence of the XMCD of the films. 

\begin{figure}
	\includegraphics[width=\linewidth]{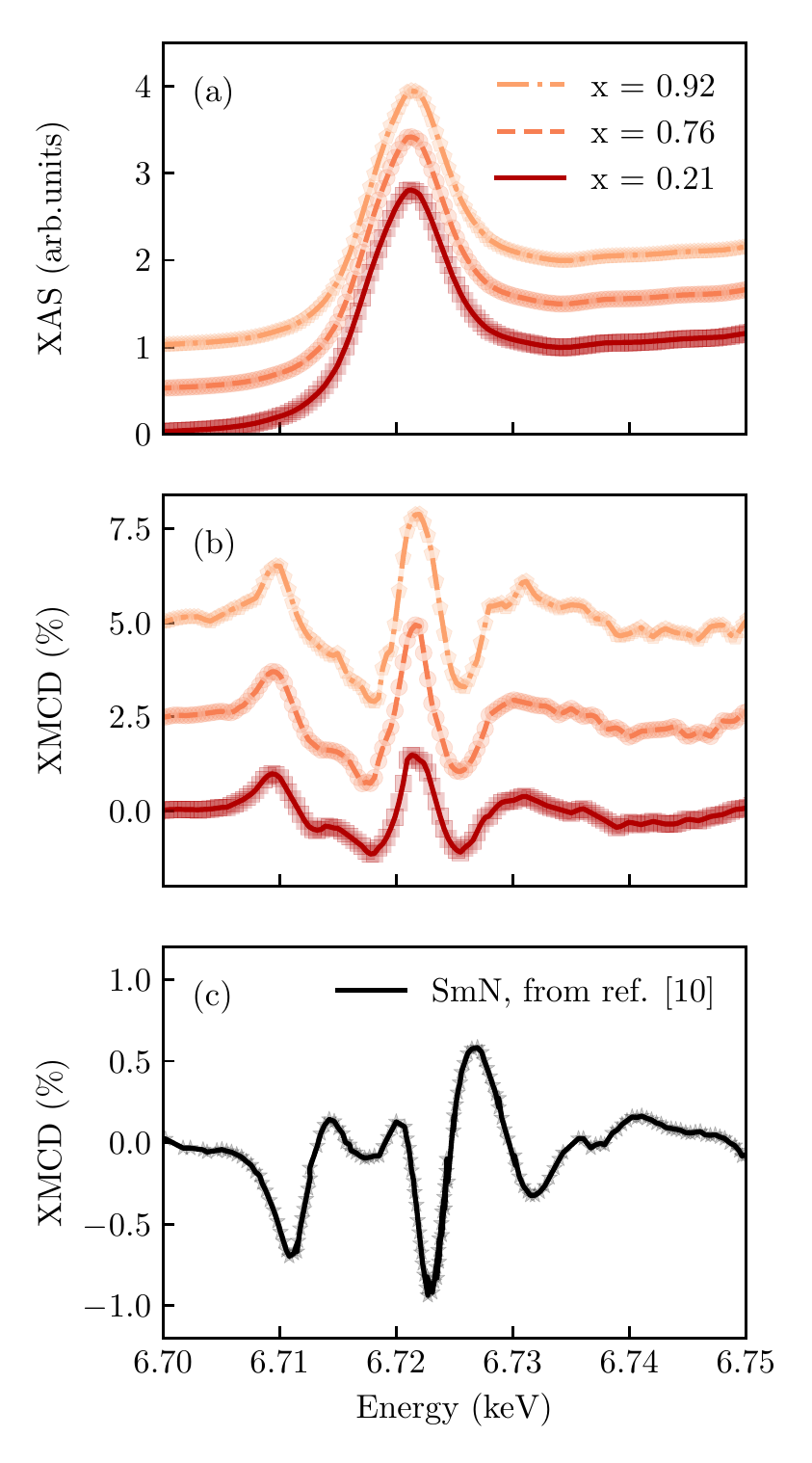}
	\caption{\small Comparison of XMCD spectra for the three Gd$_x$Sm$_{1-x}$N films at the Sm L$_{3}$ edge at $5$~K. All measurements were cooled to $5$~K in an applied field of $6$~T. Panel (a) shows the average of the x-ray absorption measurements for left- and right- circularly polarised light, normalised to the continuum edge jump. Panel (b) shows the XMCD, defined as the difference between the left- and right- circularly polarised absorptions. Panel (c) shows the XMCD Spectrum for a SmN film, data taken from~\textcite{Anton2013}.}
	\label{Alloys-SmL3}
\end{figure}

Figure \ref{Alloys-SmL3} shows Sm L$_3$ XAS and XMCD spectra for the three Gd$_{x}$Sm$_{1-x}$N films, with a spectrum taken from a homogeneous SmN film \cite{Anton2013} shown for comparison in Figure \ref{Alloys-SmL3}c. The spectra in the solid-solution films have the opposite sign to that seen for pure SmN, a clear signature of the Zeeman-exchange conflict. As discussed in the Introduction, under solely the Zeeman interaction in SmN the $4f$ \emph{spin} magnetic moment aligns in opposition to the applied field. This is reversed in the Gd$_{x}$Sm$_{1-x}$N films by Gd-Sm exchange that couples the Sm spin alignment to the strongly Zeeman-aligned Gd$^{3+}$ spins. Moreover, the magnitude of the XMCD at the Sm L$_{3}$ in Gd$_{x}$Sm$_{1-x}$N clearly increases with the value of $x$, indicating the increasing alignment of $\langle S_{z} \rangle$ in films with increasing Gd-content. 

A subtle contrast between the Gd$_x$Sm$_{1-x}$N (Figure \ref{Alloys-SmL3}b) and the SmN XMCD (Figure \ref{Alloys-SmL3}c) is an enhanced strength across the range $6.715-6.720$~keV in the present films, seen most strongly in the most Gd-rich film ($x = 0.92$). Significantly, even in SmN the weak signal at those energies was not predicted by an XMCD-based calculation~\cite{Anton2013}. It is, however, exactly the range across which there are XAS spectral lines from Sm$^{2+}$~\cite{Beaurepaire1990, Yoshikane2020, Yoshikane2021}, suggesting that their presence here could relate to a low concentration of divalent Sm. The signal is exceedingly weak, $\approx 1$~\% of the XAS signal at the normalisation point ($\approx 6.735$~keV) so it is no surprise that there appears no clear signature of this feature in the XAS spectra. In order to limit interference from the possible divalent signal in our interpretation of the Sm$^{3+}$ XMCD signal strength, for our temperature-dependent discussion below we have chosen to use the XMCD magnitude across the $6.722-6.7256$~keV transition as a measure of the spin alignment in the Sm$^{3+}$ final states.

\begin{figure}
	\includegraphics[width=\linewidth]{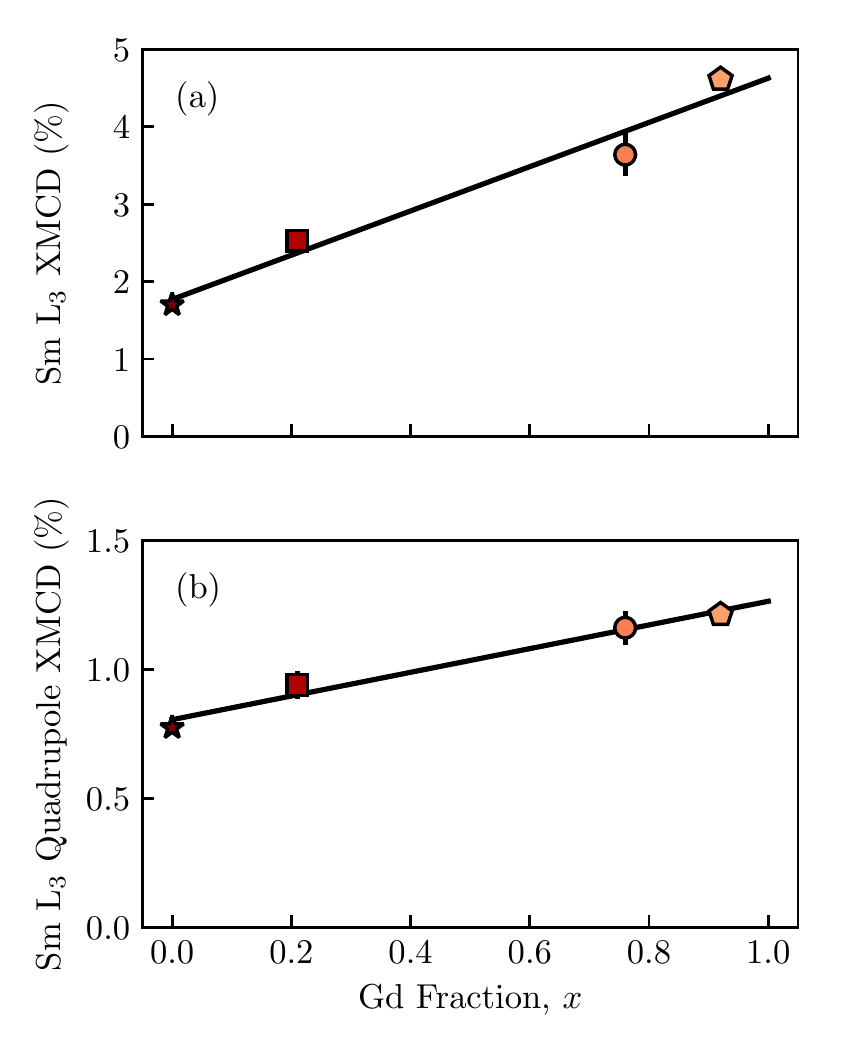}
	\caption{\small Magnitude of the Sm L$_{3}$ absorption features as a function of the Gd fraction, $x$. Panel (a) shows the magnitude for the conventional Sm$^{3+}$ absorption channel. Panel (b) shows the magnitude of the electric quadrupole absorption. The star marker represents the data taken from~\textcite{Anton2013}.}
 	\label{Sm_align_x}
\end{figure}

The feature that appears at $\approx 6.709$~keV in Figure \ref{Alloys-SmL3} is also not predicted in a calculated spectrum, and it lies at an energy that would imply an empty state deep within the valance band, well below the Fermi energy. It has been previously identified as an electric quadrupole (EQ) $2p$ to $4f$ transition~\cite{Anton2013}, in which the final $4f$ state appears, as expected within XAS, at the $4f$ energy reduced in the presence of the $2p$ core hole~\cite{Parlebas2006}. Such features have been attributed to electric quadrupole absorption in many rare-earth insulators~\cite{Neumann1999} and this attribution has been confirmed in measurements of dysprosium using angular resolved XAS~\cite{Lang1995}. It is natural to regard this feature in XMCD spectra as proportional to the spin alignment within the Sm$^{3+}$ $4f$ shell, and indeed it does strengthen across the three films, though not quite commensurate in strength with the $5d$ signals (see Figure \ref{Sm_align_x}). An analogous situation is seen also in SmN/GdN superlattices, and signals that the spin-alignment of the Sm$^{3+}$ $5d$ states is enhanced by neighboring Gd$^{3+}$ ions. The contraction of the radial $5d$ wavefunction that stems from $4f-5d$ exchange interaction leads to an enhancement of the $2p\rightarrow5d$ transition~\cite{Parlebas2006, Baudelet1993, Harmon1974}, so it is no surprise to see that the increased spin alignment of the $4f$ states is reflected with a more pronounced intensity in the electric dipole absorption channel. It is this same contraction that prevents the use of the usual XMCD sum rules to separate the orbital and spin contributions to the magnetic moment~\cite{Parlebas2006, Carra1991}.


\begin{figure}
	\includegraphics[width=\linewidth]{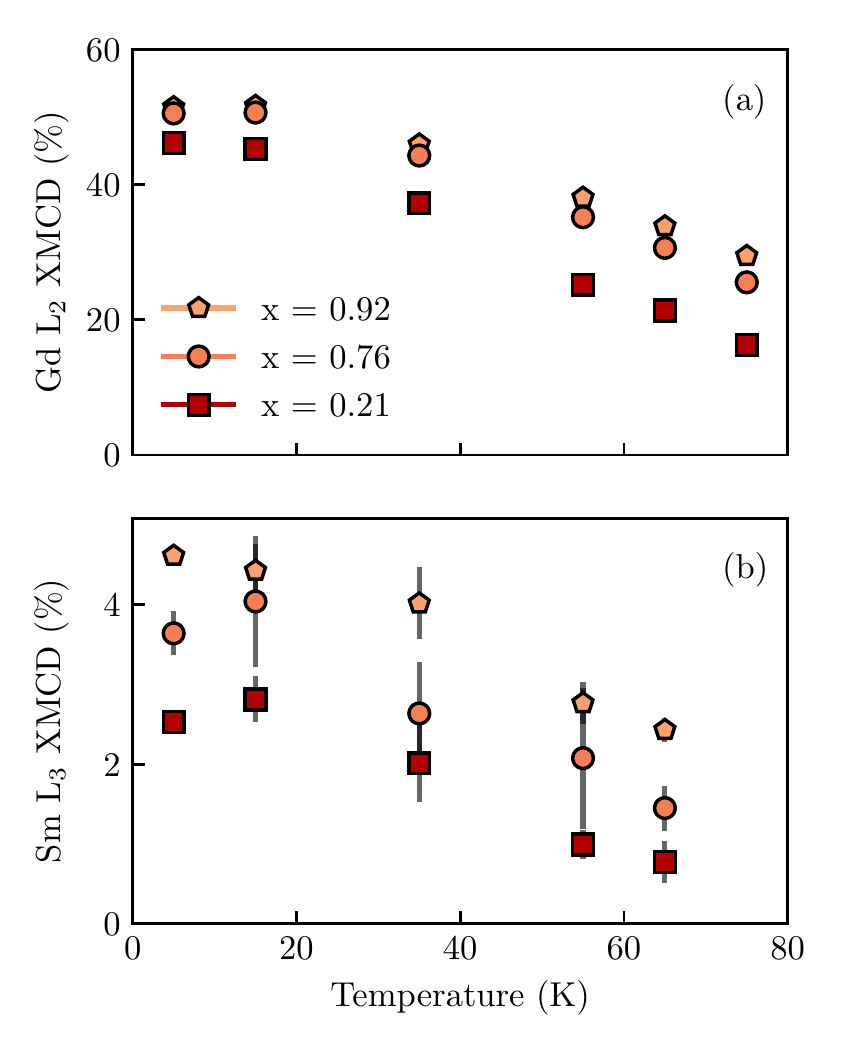}
	\caption{\small Strengths of the (a) Gd$^{3+}$ L$_{2}$-edge and (b) Sm$^{3+}$ L$_{3}$-edge features measured using XMCD as a function of temperature. Error bars are smaller than the markers where not visible.} 
	\label{Alloys-ratiovtemp}
\end{figure}


The influence of the solid-solution composition on exchange interactions is also seen in the temperature-dependent XMCD strengths, shown in Figure \ref{Alloys-ratiovtemp} as the spin alignments inferred by the Gd L$_2$ and the Sm L$_3$ edges. Looking first at Gd, there is a factor of $2$ drop of the $80$~K alignment from that at low-temperature in both of the films with high Gd concentration. Recalling that these films show a Curie temperature in the range of $60-70$~K, and that the XMCD was performed at a field of $6$~T, there is no surprise that the alignment remains strong to this temperature. In contrast, the film with only $0.21$ Gd shows a much more rapid drop. The Gd concentration for this film is within uncertainty of the site percolation threshold on the FCC lattice~\cite{Marck1997}; and this suggests the potential for a reliance in this film on the weaker Gd-Sm-Gd exchange for full ferromagnetic spin alignment on the Gd ions.The dominance of the exchange interaction over the Zeeman interaction on the Sm ions leads to a situation in which the temperature-dependence of the Sm alignment follows that of Gd ions rather closely in all films. 

\begin{figure}
	\includegraphics[width=\linewidth]{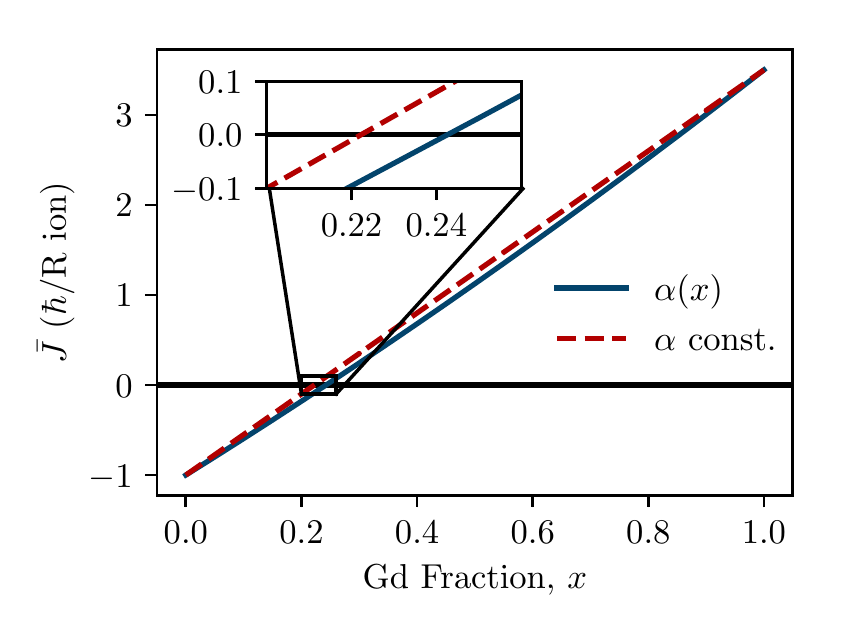}
	\caption{\small Model of the net (volume-averaged) angular momentum, $\bar{J}$, as a function of Gd fraction, $x$. The simple model is a linear interpolation of total angular momentum between $\langle J_{z} \rangle = -1$ ($\hbar$ per R ion) for SmN (where $\langle L_{z} \rangle = -2\hbar$ and $\langle S_{z} \rangle = \hbar$) and $\langle J_{z} \rangle = J = 7/2$ for GdN. Taking into account the variation of $\alpha (x)$ implied by Figure~\ref{Sm_align_x} alters the net angular momentum, shifting the compensation point.}
	\label{J_x}
\end{figure}

Returning to the results of Figure~\ref{Sm_align_x}, a consequence of the varying  Sm$^{3+}$ spin alignment is its influence on the Gd concentration of the angular momentum compensation point of Figure \ref{Schematic}(c). That compensation occurs at $x \approx 0.2$ for the $\langle S_{z} \rangle\approx\hbar$ found in SmN, but the stronger Sm$^{3+}$ alignment in these solid solutions shifts the compensation point to higher Gd concentration. A quantitative interpretation of the shift relies on a measure of the enhancement of the $4f$ spin-alignment $\langle S_z(x)\rangle$, for which we rely on the increase of the $4f$ quadrupole feature at $\approx~6.709$~keV shown in Figure \ref{Sm_align_x}. Those data imply that 
the magnitude of the Sm$^{3+}$ electric quadrupole feature is enhanced by $\sim 1.5$ between $x=0$ and $x \rightarrow 1$. The strong spin-orbit interaction ensures the Sm$^{3+}$ net angular momentum (which is negative with respect to the spin direction, due to the orbital-dominant magnetism) will also increase linearly with $x$, leading ultimately to the shift of the compensation point by $\approx 0.02$ as shown in Figure \ref{J_x}.  

\section{Conclusions}

We have probed the alignment of the moments in Gd$_{x}$Sm$_{1-x}$N by L-edge XMCD tracing out the spin alignments $\langle S_{z} \rangle$ independently on the Sm$^{3+}$ and Gd$^{3+}$ ions. Special care was taken to follow the XMCD signal amplitudes and the spin alignments $\langle S_{z} \rangle$ of the two ionic species as a function of the composition across the GdN/SmN solid solutions. 

The XMCD study was supported by thorough magnetisation studies of the films. It is noted that the magnetisation is overwhelmingly dominated by the Gd$^{3+}$ spin alignment, since the magnetic moment on the Sm$^{3+}$ is more than two orders of magnitude the smaller. Magnetic measurements revealed the temperature-dependent magnetisation as well and complete hysteresis curves in the ferromagnetic phase that were used to inform XMCD measurements and limit those to magnetic fields adequate to ensure saturated magnetisation. There was excellent agreement between the magnetisation and Gd $\langle S_{z} \rangle$, but it is only XMCD that is capable of revealing the spin alignment $\langle S_{z} \rangle$ on the Sm$^{3+}$ ions, the primary outcome of the study. 

The sign of the Sm XMCD, opposite to that in SmN, immediately identifies that the alignment of the Sm and Gd spins by the Sm-Gd exchange interaction dominates the Sm$^{3+}$ Zeeman interaction, which on its own acts to align the \emph{orbital} magnetic moment in SmN~\cite{Anton2013, McNulty2015}. The data further show that the exchange dominance strengthens the Sm$^{3+}$ spin alignment in the presence of Gd$^{3+}$ ions; the projection $\langle S_{z} \rangle\approx\hbar$ in SmN, rising to $\langle S_{z} \rangle\approx3\hbar/2$ for isolated Sm$^{3+}$ ions in a predominantly GdN host.  The composition-dependent alignment specifies that the composition at which an angular-momentum compensation can be expected is at higher values of $x$ than implied by measurements of pure SmN films. 




~\begin{acknowledgments}
The Synchrotron Radiation experiments were performed at the BL39XU beamline of SPring-8 with the approval of the Japan Synchrotron Radiation Research Institute (JASRI), Proposal No: 2020A1250. The MacDiarmid Institute is supported under the New Zealand Centres of Research Excellence Programme.
\end{acknowledgments}

\bibliography{PaperXMCD}

\end{document}